\documentclass[twocolumn]{aastex62}

\usepackage{graphicx}
\usepackage{amssymb}
\usepackage{amsmath}
\usepackage{natbib}
\usepackage{ulem}
\accepted{June 11, 2019}

\shorttitle{Signatures of stellar accretion in MaNGA early-type galaxies}
\shortauthors{Oyarz\'un et al.}

\begin{document}
	
	\title{Signatures of stellar accretion in MaNGA early-type galaxies}

	\email{goyarzun@ucsc.edu}
	
	\author{Grecco A. Oyarz\'un}
	\affiliation{Astronomy Department, University of California, Santa Cruz, CA 95064, USA}
	
	\author{Kevin Bundy}
	\affiliation{University of California Observatories - Lick Observatory, University of California, Santa Cruz, CA 95064, USA}
	
	\author{Kyle B. Westfall}
	\affiliation{University of California Observatories - Lick Observatory, University of California, Santa Cruz, CA 95064, USA}
	
	\author{Francesco Belfiore}
	\affiliation{University of California Observatories - Lick Observatory, University of California, Santa Cruz, CA 95064, USA}
	\affiliation{European Southern Observatory, Karl-Schwarzschild-Str. 2, Garching bei M\"unchen, 85748, Germany}
	
	\author{Daniel Thomas}
	\affiliation{Institute of Cosmology and Gravitation, University of Portsmouth, Burnaby Road, Portsmouth, PO1 3FX, UK}
	
	\author{Claudia Maraston}
	\affiliation{Institute of Cosmology and Gravitation, University of Portsmouth, Burnaby Road, Portsmouth, PO1 3FX, UK}	

	\author{Jianhui Lian}
	\affiliation{Institute of Cosmology and Gravitation, University of Portsmouth, Burnaby Road, Portsmouth, PO1 3FX, UK}
	
	\author{Alfonso Arag\'on-Salamanca}
	\affiliation{School of Physics and Astronomy, University of Nottingham, University Park, Nottingham, NG7 2RD, UK}
	
	\author{Zheng Zheng}
	\affiliation{National Astronomical Observatories, Chinese Academy of Sciences, A20 Datun Road, Beijing, China}
	\affiliation{CAS Key Laboratory of FAST, NAOC, Chinese Academy of Sciences}
	
	\author{Violeta Gonzalez-Perez}
	\affiliation{Institute of Cosmology and Gravitation, University of Portsmouth, Burnaby Road, Portsmouth, PO1 3FX, UK}
	\affiliation{Energy Lancaster, Lancaster University, Lancaster LA14YB, UK}
	
	\author{David R. Law}
	\affiliation{Space Telescope Science Institute, 3700 San Martin Drive, Baltimore, MD 21218, USA}
	
	\author{Niv Drory}
	\affiliation{McDonald Observatory, The University of Texas at Austin, 1 University Station, Austin, TX 78712, USA}
	
	\author{Brett H. Andrews}
	\affiliation{PITT PACC, Department of Physics and Astronomy, University of Pittsburgh, Pittsburgh, PA 15260, USA}
	
	
	
	
	\begin{abstract}

		The late assembly of massive galaxies is thought to be dominated by stellar accretion in their outskirts (beyond 2 effective radii R$_{\mbox{\scriptsize e}}$) due to dry, minor galaxy mergers. We use observations of 1010 passive early-type galaxies (ETGs) within $z<0.15$ from SDSS IV MaNGA to search for evidence of this accretion. The outputs from the stellar population fitting codes FIREFLY, pPXF, and Prospector are compared to control for systematic errors in stellar metallicity (Z) estimation. We find that the average radial logZ/Z$_{\odot}$ profiles of ETGs in various stellar mass (M$_*$) bins are not linear. As a result, these profiles are poorly characterized by a single gradient value, explaining why weak trends reported in previous work can be difficult to interpret. Instead, we examine the full radial extent of stellar metallicity profiles and find them to flatten in the outskirts of M$_*\gtrsim 10^{11}$M$_{\odot}$ ETGs. This is a signature of stellar accretion. Based on a toy model for stellar metallicity profiles, we infer the \textit{ex-situ} stellar mass fraction in ETGs as a function of M$_*$ and galactocentric radius. We find that \textit{ex-situ} stars at R$\sim$2R$_{\mbox{\scriptsize e}}$ make up 20\% of the projected stellar mass of M$_*\lesssim 10^{10.5}$M$_{\odot}$ ETGs, rising up to 80\% for M$_*\gtrsim 10^{11.5}$M$_{\odot}$ ETGs.
		
	\end{abstract}
	
	
	
	
	
	\section{Introduction}
	\label{1}
	
	The effective radii (R$_{\mbox{\scriptsize e}}$) of $z\sim 0$ early-type galaxies (ETGs) are observed to be a factor of three to six larger than those of their $z\sim 2$ counterparts (\citealt{toft2007,cimatti2008,buitrago2008,vandokkum2010}). On the other hand, the stellar masses (M$_*$) of local ETGs have only increased by a factor of two since $z\sim 2$ (\citealt{daddi2005,trujillo2006b,trujillo2006a,trujillo2007,zirm2007,vanderwel2008,vandokkum2008,damjanov2009,cassata2010,cassata2011}). While galaxies quenched at later times tend to be larger, driving the average R$_{\mbox{\scriptsize e}}$ upward (progenitor bias; e.g. \citealt{valentinuzzi2010,carollo2013}), this alone is not sufficient to explain size growth (e.g. \citealt{furlong2017}). Late stellar accretion in spheroidal, or even disk configurations (\citealt{graham2015}), appears to be required, especially at the high M$_*$ end (M$_*>$10$^{10.5}$M$_{\odot}$, e.g. \citealt{genel2018}). Minor mergers have been shown to be particularly efficient at increasing the R$_{\mbox{\scriptsize e}}$ of ETGs while keeping their M$_*$ roughly constant (e.g. \citealt{bezanson2009,hopkins2010,barro2013,cappellari2013b,wellons2015}).  
	
	These ideas are at the basis of the current cosmological picture for structure evolution at $z<2$, in which massive systems accrete stellar envelopes from satellite galaxies (\citealt{oser2010,oser2012,johansson2012,moster2012,furlong2017}). In this framework, stars that formed within their host galaxies tend to dominate at the center, whereas accreted stars begin to do so in the outskirts (R$\sim$2R$_{\mbox{\scriptsize e}}$; \citealt{rodriguez-gomez2016}) and in the lower surface brightness regions beyond 2R$_{\mbox{\scriptsize e}}$ known as stellar halos (\citealt{zolotov2009,tissera2013,tissera2014,cooper2015}). These stellar populations of different origin are usually referred to as \textit{in-situ} and \textit{ex-situ}, respectively. Several simulations have made predictions about observational signatures of the predicted radial transition from \textit{in-situ} to \textit{ex-situ} (e.g. \citealt{pillepich2014}). Among stellar population tracers, stellar metallicity is expected to be one of the most sensitive to this transition (e.g. \citealt{cook2016}).
	
	In the absence of late-time minor mergers, the radial stellar metallicity profiles are predicted to be negative (\citealt{kobayashi2004,pipino2010,taylor2017}). This implies that the outer parts of ETGs tend to be more metal-poor than the inner parts. Albeit with significant variance (\citealt{lackner2012,hirschmann2015}), the deposition of accreted stars in the outskirts of galaxies induces flattening of the \textit{in-situ} profile (\citealt{cook2016,taylor2017}). Since mergers are expected to have a larger effect on more massive systems, the resulting prediction is that the stellar metallicity profiles of ETGs are flatter toward higher M$_*$, especially in the stellar halos (\citealt{cook2016}). 
	
	These theoretical predictions have motivated the search for observational signatures of stellar accretion. Using long-slit spectroscopy, \citealt{carollo1993} estimated the strength of metallicity-sensitive stellar absorption features as a function of galactocentric radius in 42 nearby galaxies. Though larger samples can be studied using photometric surveys (e.g. \citealt{labarbera2005,labarbera2011,tortora2010,tortora-napolitano2012}), spectroscopy is critical for breaking the age-metallicity degeneracy. More recently, studies of stellar populations in nearby galaxies have benefited from integral field unit (IFU) surveys like MASSIVE (\citealt{greene2013,greene2015}), CALIFA (\citealt{sanchez2012}), SAMI (\citealt{allen2015}), and MaNGA (Mapping Nearby Galaxies at Apache Point Observatory; \citealt{bundy2015}). In particular, MaNGA observations extend to the outskirts of galaxies (beyond 2 R$_{\mbox{\scriptsize e}}$), starting to probe the radii at which the signatures of minor mergers are predicted to appear (e.g. \citealt{cook2016}).
	
	Stellar metallicity profiles are typically characterized by radial gradients, estimated by fitting a linear form to the profile between the center and 1-2R$_{\mbox{\scriptsize e}}$ (e.g. \citealt{zheng2017,goddard2017b,li2018}). In agreement with simulations, the metallicity gradients of ETGs tend to be negative (e.g. \citealt{rawle2010,gonzalez-delgado2015,roig2015}). However, the dependence of the gradient slope on M$_*$ remains unclear. Based on a sample of $\sim10^3$ galaxies from the MaNGA survey, \citet{zheng2017} find weak or no correlation between the gradients and M$_*$. Using data from the same survey, \citet{goddard2017a} find that gradients are steeper with increasing M$_*$, although with low significance. Though also based on MaNGA, \citet{li2018} find shallower gradients at higher central velocity dispersions ($\sigma_*>100$km/s). There are several possible sources for these discrepancies, from stellar population synthesis approach (see \citealt{conroy2013}) to fitting method. Another important factor, as we show in this paper, is that the stellar metallicity profiles of ETGs are not well described by a linear fit.
	
	In this work, we examine the full radial extent of metallicity profiles from spatially resolved spectroscopy of 1010 ETGs from MaNGA. We inform our interpretation of the stellar metallicity profiles by using results from hydrodynamical simulations (e.g. \citealt{cook2016,rodriguez-gomez2016,dsouza-bell2018}). This paper is structured as follows. In Section \ref{2} we define our sample. In Section \ref{3}, we describe the stellar population fitting process with the codes FIREFLY (\citealt{wilkinson2017}), pPXF (\citealt{cappellari-emsellem2004,cappellari2017}), and Prospector (\citealt{leja2017}). We show our results in Section \ref{4} and discuss the implications in Section \ref{5}. We summarize in Section \ref{6}. This work adopts H$_0=70$ km s$^{-1}$Mpc$^{-1}$ and all magnitudes are reported in the AB system (\citealt{oke1983}).

	\begin{figure*}
		\centering
		\includegraphics[width=7in]{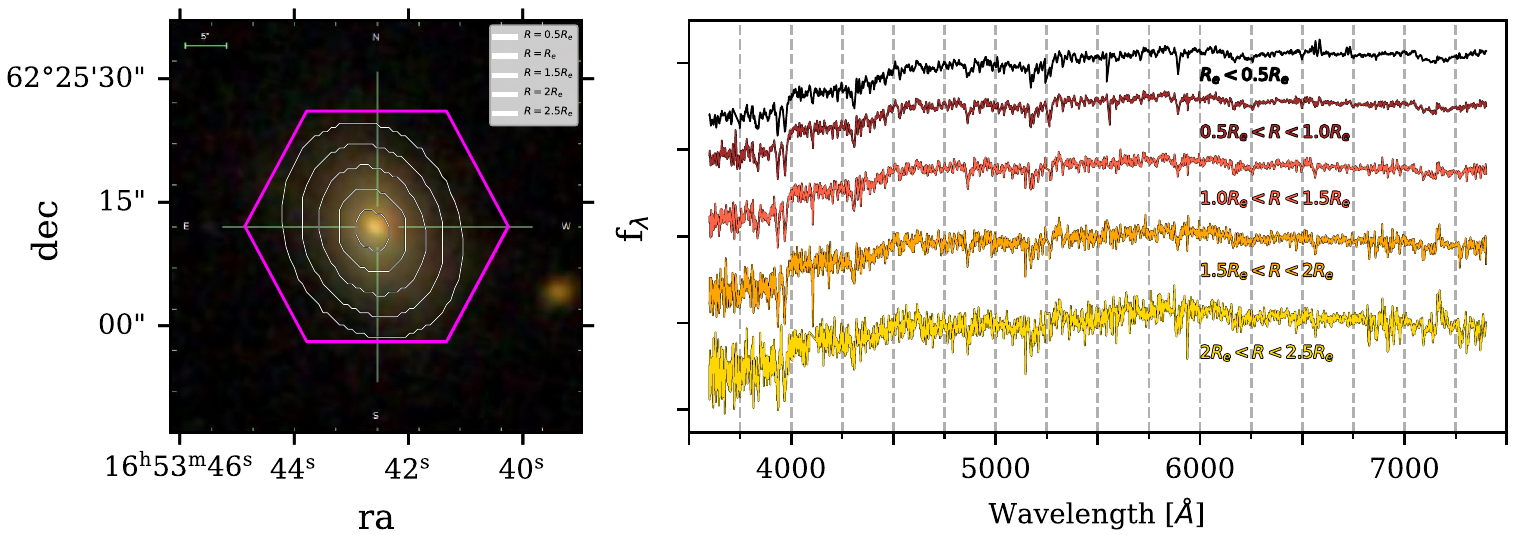}
		\caption{Illustration of our analysis on MaNGA galaxy 1-22298, one of 1010 ETGs in our sample. Left: SDSS r-band image. The MaNGA IFU footprint is overlaid in magenta. We also show in white the five annuli defined for this galaxy. Right: Co-added spectra for every annulus from the center to the outskirts.}
		\label{fig1}
	\end{figure*}
	
	\section{Dataset}
	\label{2}
	
	The MaNGA survey (\citealt{bundy2015,yan2016b}) is part of the fourth generation of SDSS (\citealt{york2000,gunn2006,blanton2017}), and is on track to provide spatially resolved spectra for ten thousand nearby galaxies ($z<0.15$) by the end of 2020. By means of integral field unit spectroscopy (IFS; \citealt{smee2013,drory2015,law2015}), every galaxy is observed with 19-to-127 fiber bundles with diameters varying between 12\farcs 5 and 32\farcs 5. The resulting radial coverage reaches between 1.5R$_{\mbox{\scriptsize e}}$ and 2.5R$_{\mbox{\scriptsize e}}$ for most targets (\citealt{wake2017}; see Figure \ref{fig1}). The spectra cover the wavelength range 3600-10300 \AA \ at a resolution of R$\sim$2000.
	
	All MaNGA data used in this work were reduced by the Data Reduction Pipeline (DRP; \citealt{law2016,yan2016a}). The reduced spectra have a median spectral resolution of $\sigma$=72 km s$^{-1}$. The data cubes typically reach a 10$\sigma$ continuum surface brightness of $\mu$=23.5 mag arcsec$^{-2}$, and their astrometry is measured to be accurate to 0\farcs1 (\citealt{law2016}). De-projected distances and stellar kinematic maps have been calculated by the MaNGA Data Analysis Pipeline (DAP; \citealt{westfall2019}). This work also makes use of Marvin (\citealt{cherinka2017}), the specially designed tool for access and handling of MaNGA data\footnote{\href{https://api.sdss.org/doc/manga/marvin}{https://api.sdss.org/doc/manga/marvin}}.
	
	This paper is based on the SDSS Data Release 15 (DR15), which consists of the observations of the first 4675 MaNGA targets. We extracted the stellar masses (M$_*$), Sersic indices ($n_{\mbox{\scriptsize Sersic}}$), and effective radii (R$_{\mbox{\scriptsize e}}$) of these galaxies from the publicly available NASA-Sloan Atlas\footnote{\href{http://nsatlas.org}{http://nsatlas.org}}(NSA). In particular, the M$_*$ estimates were derived using a k-correction fit to the Sersic fluxes (\citealt{blanton2007}), adopting the \citet{bruzualcharlot2003} stellar population models and a \citet{chabrier2003} initial mass function (IMF). They also assumed H$_0=100$ km s$^{-1}$Mpc$^{-1}$, but we scaled them for an H$_0=70$ km s$^{-1}$Mpc$^{-1}$ cosmology. The $n_{\mbox{\scriptsize Sersic}}$ estimates were obtained from one-component, two-dimensional fits to r-band images. The R$_{\mbox{\scriptsize e}}$ are determined using an elliptical Petrosian analysis of the $r$-band image from the NSA. All NSA measurements use the detection and deblending technique described in \citet{blanton2011}.

	To select ETGs, we first applied the morphological cut $n_{\mbox{\scriptsize Sersic}}>2.5$ (e.g. \citealt{blanton2003,blanton2005,peng2010}). In addition, we selected passive ETGs by using the average H$\alpha$ equivalent width across the galaxy -EW(H$\alpha)$- as proxy for specific star-formation rate (sSFR). The cut was EW(H$\alpha)<3$ \AA, which is commonly used to distinguish between ionization due to smooth background of hot evolved stars and due to star formation and AGN (\citealt{cidfernandes2011}; see also \citealt{belfiore2016}). This yielded a sample with 1101 galaxies. We also limited the central velocity dispersions and stellar masses of our sample to the ranges $\sigma_*<400$ km s$^{-1}$ and $10<$logM/M$_*<12$, respectively. We performed these cuts to provide a relatively uniform distribution of ETGs over M$_*$. The final outcome was a sample of 1010 ETGs. We did not remove quiescent galaxies with significant stellar disks from the sample. From visual inspection, we estimate the fraction of lenticulars (S0s) to be $\lesssim 20$ \%. However, we acknowledge the challenge of achieving precise S0 classification of SDSS galaxies (see \citealt{nair2010}). Our selection may also miss blue ellipticals, but their number fraction is $\lesssim 5\%$ for our M$_*$ range (\citealt{kannappan2009}). Our goal here is to study a generally passive sample of spheroidal galaxies. We delay to future work a characterization of stellar populations in more finely discriminated morphological types.
	
	\section{Methodology}
	\label{3}
	
	\subsection{Radial  binning}
	\label{3.1}
	Using the R$_{\mbox{\scriptsize e}}$ value of every galaxy, we associated elliptical polar radii to all spaxels in units of R$_{\mbox{\scriptsize e}}$. These account for the axis ratio of every object, which were measured on the r-band photometry. We then binned them into the five annuli R/R$_{\mbox{\scriptsize e}}$= $[0, 0.5]$, $[0.5, 1]$, $[1, 1.5]$, $[1.5, 2]$, and $[2, 2.5]$. This is shown for a sample galaxy on the left panel of Figure \ref{fig1}.
	
	After binning, we shifted every spectrum back to the rest-frame using the stellar systemic velocity ($v_*$) maps calculated by the DAP. We used the maps computed with a Voronoi binning scheme that aims for a minimum signal-to-noise ratio of 10 per bin. For each galaxy, we co-added the spectra in every annular bin. We did not convolve the spectra to a common $\sigma_*$ prior to stacking. After co-addition, we ran pPXF (\citealt{cappellari-emsellem2004,cappellari2017}) with the MILES Single Stellar Population (SSP) library (\citealt{vazdekis2010}) on the stacked spectra to measure the co-added $v_*$ and $\sigma_*$. The right panel of Figure \ref{fig1} shows the five co-added spectra for a sample galaxy.
	
	

	\subsection{Stellar population fitting}
	
	Estimates of stellar population parameters like stellar metallicity can be obtained by full spectral fitting, but depend sensitively on the adopted priors, assumptions used to generate template spectra (\citealt{conroy2013}), and fitting method. To mitigate the effect of systematic biases from any one approach, we applied three independent codes to the same data and examine the differences that arise.
	
	The first code we ran was the public version of FIREFLY\footnote{\href{http://www.icg.port.ac.uk/FIREFLY/}{FIREFLY - A full spectral fitting code \\ http://www.icg.port.ac.uk/FIREFLY/}}\footnote{\href{https://github.com/FireflySpectra/firefly\_release}{https://github.com/FireflySpectra/firefly\_release}}(\citealt{comparat2017,wilkinson2017,goddard2017a}). This $\chi^2$ minimization code decouples stellar populations from dust by removing the low-order continuum shape before performing the model fitting. Hence, it focuses on high frequency modes in the spectra to infer stellar ages and metallicities. SSPs of different ages and metallicities are added iteratively until the improvement in $\chi^2$ is negligible.
	
	We ran the code with the stellar population models of \citet{maraston-stromback2011}, MILES stellar library (\citealt{sanchez-blazquez2006}), and Chabrier IMF (\citealt{chabrier2003}). We used a set of SSPs covering an age grid between 6.5 Myr and 15 Gyr, while the sampled stellar metallicities were logZ/Z$_\odot$=-2.3, -1.3, -0.3, 0.0, and 0.3. The library spans the wavelength range 4000\AA \ to 7400\AA. As shown in \citealt{wilkinson2017}, FIREFLY effectively recovers stellar population parameters for spectra with S/N$>10$ (see also \citealt{goddard2017b}). To limit the systematics in the measurements from Firefly, we excluded any co-added spectra with S/N$<10$. We also masked emission lines. Fitting with FIREFLY took, on average, a minute per spectrum on a single core. Throughout this paper, we show light-weighted measurements, although we find similar results when using the mass-weighted counterparts.
		
	We also ran pPXF\footnote{\href{https://www-astro.physics.ox.ac.uk/~mxc/software/\#ppxf}{pPXF \\ https://www-astro.physics.ox.ac.uk/~mxc/software/\#ppxf}}(\citealt{cappellari-emsellem2004,cappellari2017}) on our spectra. This code applies a penalized maximum likelihood approach to fit libraries of stellar population templates to observed data. Since this code penalizes pixels that are not well characterized by the templates, it minimizes template mismatch. We ran it with the included library of SSPs based on the MILES stellar library (\citealt{sanchez-blazquez2006,vazdekis2010}).
		
	We simultaneously fitted for the gas and the stars, allowing for two moments in gas kinematics and four in stellar kinematics. We chose not to smooth the distribution of template weights (i.e., no regularization). After the best linear combination of templates was found, we added several realizations of the noise in the spectra to the best fit. This allowed us to characterize the uncertainties in the reported stellar population parameters. On average, our runs of pPXF took about a minute per spectrum on a single core.
	
	The third stellar population fitting code we ran was Prospector\footnote{\href{https://github.com/bd-j/prospector/blob/master/doc/index.rst}{Prospector \\ https://github.com/bd-j/prospector/blob/master/doc/index.rst}}(\citealt{leja2017}). This code is based on the stellar population synthesis code FSPS\footnote{\href{https://github.com/cconroy20/fsps}{FSPS: Flexible Stellar Population Synthesis \\ https://github.com/cconroy20/fsps}}(\citealt{conroy2009,conroy-gunn2010}), which generates composite stellar spectra for a variety of prescriptions for stellar population synthesis and evolution. This allows Prospector to sample the posterior distribution of a user-defined parameter space, while formally characterizing uncertainties and degeneracies. We chose the MILES stellar library (\citealt{sanchez-blazquez2006}), MIST isochrones (\citealt{dotter2016,choi2016}), and Kroupa IMF (\citealt{kroupa2001}) as inputs. We also masked emission lines prior to fitting.
		
	Since we fitted old stellar populations, we modeled the spectra with exponentially decaying ($\tau$) star-formation histories to speed up the fitting process. In addition to $\tau$, our parameter space included the optical depth of dust in the V-band, and stellar ages, metallicities, masses, and velocity dispersions. Our priors are shown in Table \ref{table1}. To derive the posterior distributions, we used the Dynamic Nested Sampling package dynesty\footnote{\href{https://github.com/joshspeagle/dynesty/blob/master/docs/source/index.rst}{dynesty \\ https://github.com/joshspeagle/dynesty/blob/master/docs \\ /source/index.rst}}(\citealt{speagle2019}). On average, convergence of Prospector with dynesty was achieved after an hour per spectrum on a single core.

	\begin{deluxetable}{c|c}
		\centering
		\tablecaption{Priors used in our Prospector runs}
		\tabletypesize{\footnotesize}
		\tablewidth{0pt}
		\tablehead{
			Parameter & Prior}
		\startdata
		\label{table1}			
		$\tau$ & $\mbox{LogUniform}(10^{-2}, 10)$ \\
		$\mbox{dust2}$ & $\mbox{TopHat}(0, 1)$ \\
		$\mbox{Stellar age [Gyr]}$ & $\mbox{TopHat}(5, 14)$ \\
		$\mbox{Stellar metallicity [logZ/Z$_{\odot}$]}$ & $\mbox{TopHat}(-2, 0.3)$ \\
		$\mbox{Stellar mass [M$_{\odot}$]}$ & $\mbox{LogUniform}(10^5, 10^{12})$ \\
		$\mbox{$\sigma_*$ [km/s]}$ & $ \mbox{TopHat}(10, 400)$										
		\enddata	
	\end{deluxetable}

\begin{deluxetable*}{c|c|c|c|c|c}
	\centering
	\tablecaption{Number of spectra used in our analysis\tablenotemark{a,b}}
	\tabletypesize{\footnotesize}
	\tablewidth{0pt}
	\tablehead{
		Stellar mass bin [M$_{\odot}$]& R$<$0.5R$_{\mbox{\scriptsize e}}$ & 0.5R$_{\mbox{\scriptsize e}}$$<$R$<$1R$_{\mbox{\scriptsize e}}$ & 1R$_{\mbox{\scriptsize e}}$$<$R$<$1.5R$_{\mbox{\scriptsize e}}$ &  1.5R$_{\mbox{\scriptsize e}}$$<$R$<$2R$_{\mbox{\scriptsize e}}$ & 2R$_{\mbox{\scriptsize e}}$$<$R$<$2.5R$_{\mbox{\scriptsize e}}$}
	\startdata
	\label{table2}
	$10^{10}-10^{10.5}$ & 174 (174)& 174 (174) & 174 (174) & 170 (170)& 167 (160)\\	
	$10^{10.5}-10^{11}$ & 267 (267)& 267 (267) & 266 (265)& 264 (264)& 252 (246)\\	
	$10^{11}-10^{11.5}$ & 420 (420) & 420 (420)& 417 (417)& 393 (392)& 347 (319)\\	
	$10^{11.5}-10^{12}$ & 148 (148)& 147 (147)& 142 (142)& 114 (111) & 88 (67)
	\enddata
	\tablenotetext{a}{Applies to pPXF and Prospector. Firefly numbers are in parenthesis (S/N$>10$ cut).}
	\tablenotetext{a}{The decrease in number of spectra with radius is a consequence of IFU coverage and quality cuts on the fits to stellar kinematics.}	
\end{deluxetable*}
	
	\section{Results}
	\label{4}
	
\begin{figure*}
		\centering
		\includegraphics[width=7in]{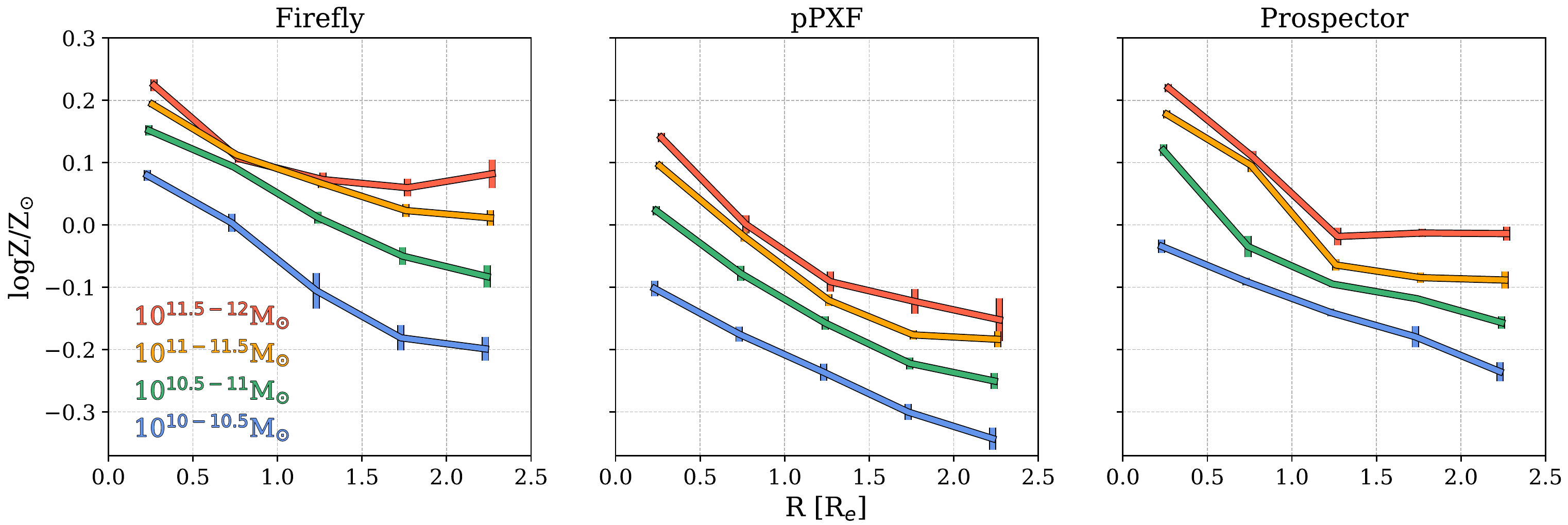}
		\caption{Median radial metallicity profiles of ETGs for different M$_*$ bins. The three panels show the profiles derived by the codes Firefly, pPXF, and Prospector. The profiles of lower mass ETGs fall linearly with galactocentric radius. As galaxy mass increases, the profiles flatten at R$>$1.5R$_{\mbox{\scriptsize e}}$.}
		\label{fig7}
	\end{figure*}

	Using the three codes described above, we derived stellar population parameters in each radial bin for all galaxies in the sample. After binning in M$_*$ (with numbers in Table \ref{table2}), we computed the average stellar metallicity profiles as a function of M$_*$ and show them in Figure \ref{fig7}. The three panels show the results from the three fitting codes. While the metallicity profiles differ in normalization and in their detailed shapes, qualitative trends are similar across the codes.
		
	We start by discussing the two notable discrepancies among the outputs. First, pPXF systematically measures metallicities $\sim 0.1$ dex lower than Firefly and Prospector. This overall offset does not correlate with S/N or M$_*$ and will not affect our primary conclusions, which are based on the shape of derived metallicity profiles. Second, Firefly outputs tend to avoid metallicities in the range logZ/Z$_\odot = [-1.3, -0.3]$, preferring higher values. This is presumably due to sampling in the stellar metallicity grid (see \citealt{wilkinson2017}). As we show in Figure \ref{fig7}, flattening of the Firefly metallicity profiles occurs at higher metallicities as a result.
	
	In nearly all radial bins, more massive galaxies exhibit more metal-rich stars. The logZ/Z$_{\odot}$ profiles of ETGs fall linearly with galactocentric radius out to 1.5R$_{\mbox{\scriptsize e}}$. Remarkably, the profiles flatten at the largest radii for M$_*>10^{11}$M$_{\odot}$. The flattening is present in the output of all three codes. Comparing a given set of profiles as a function of M$_*$, we see that the radius at which this flattening occurs moves inward as M$_*$ increases. These results are also apparent in the behavior of Lick indices Fe4531, Mgb(5178), and Fe5270 (Appendix \ref{appendix1}). The observed flattening is consistent with the signatures of stellar accretion predicted by hydrodynamical simulations (e.g. \citealt{cook2016}) and motivates the interpretative framework we discuss in Section \ref{5}.
	
	We note that even though the M$_*$ dependence of the stellar metallicity profiles is consistent across codes, the same cannot be said about the stellar age profiles (not shown). This is not surprising, since it is extremely difficult to determine the ages of stellar populations older than 9 Gyr because of the slow isochrone evolution at late times (\citealt{conroy2013}). Since radial gradients in stellar age are not predicted to capture much information about the accretion history of ETGs (\citealt{cook2016}), we leave a more detailed analysis of stellar ages for future work.
	
	Some galaxies only satisfied our quality criteria (see Section \ref{3}) at some annuli. Hence, some galaxies contributed only to some regions in the profiles of Figure \ref{fig7}. To ensure our results are not biased, as a result we constructed a subset of 822 ETGs composed only of high quality spectra (S/N$>10$ for all radii). Our results were also recovered with this subset.
	
	We have also attempted to reproduce our results using the publicly available Firefly\footnote{\href{http://www.sdss.org/dr14/manga/manga-data/manga-FIREFLY-value-added-catalog/}{MaNGA FIREFLY Value Added Catalog http://www.sdss.org/dr14/manga/manga-data/manga- FIREFLY-value-added-catalog/}} (\citealt{goddard2017a,goddard2017b}) and Pipe3D\footnote{\href{https://www.sdss.org/dr14/manga/manga-data/manga-pipe3d-value-added-catalog/}{Pipe3D Value Added Catalog  https://www.sdss.org/dr14/manga/manga-data/manga-pipe3d-value-added-catalog/}} (\citealt{sanchez2016,sanchez2018}) Value Added Catalogs, which provide spatially-resolved maps of stellar population properties for MaNGA galaxies. Unfortunately, Voronoi bins with S/N$<$10 dominate in the outermost low-surface brightness regions. Various tests have shown that stellar population codes are biased at S/N$<$10 (e.g. \citealt{wilkinson2017}). As a result of these complications, we refrained from incorporating these catalogs in our analysis.
    
	
	\section{Discussion}
	\label{5}
	
	\subsection{On the radial metallicity profiles of ETGs}
	\label{5.1}	
	
	
	In the R$<$R$_{\mbox{\scriptsize e}}$ region, \citet{martin-navarro2018} found that the stellar metallicity profiles of ETGs fall more steeply at higher $\sigma_*$ and M$_*$. Similarly, \citet{goddard2017a} reported weak evidence for a steepening of their radial gradients with M$_*$. On the other hand, \citet{kuntschner2010,tortora2010,kuntschner2015,li2018} found gradients to flatten at higher $\sigma_*$. \citet{gonzalez-delgado2015,zheng2017} claimed no clear correlation between their stellar metallicity gradients and M$_*$. Similarly, \citet{greene2013,greene2015} found no strong correlations between the shape of element abundance profiles and $\sigma_*$. In this work, we found the profiles to flatten in the outskirts for logM$_*$/M$_{\odot}\gtrsim11$. Here, we demonstrate how some of the apparent disagreement among observations may owe to the definition of metallicity gradients.
	
\begin{figure*}
		\centering
		\includegraphics[width=7in]{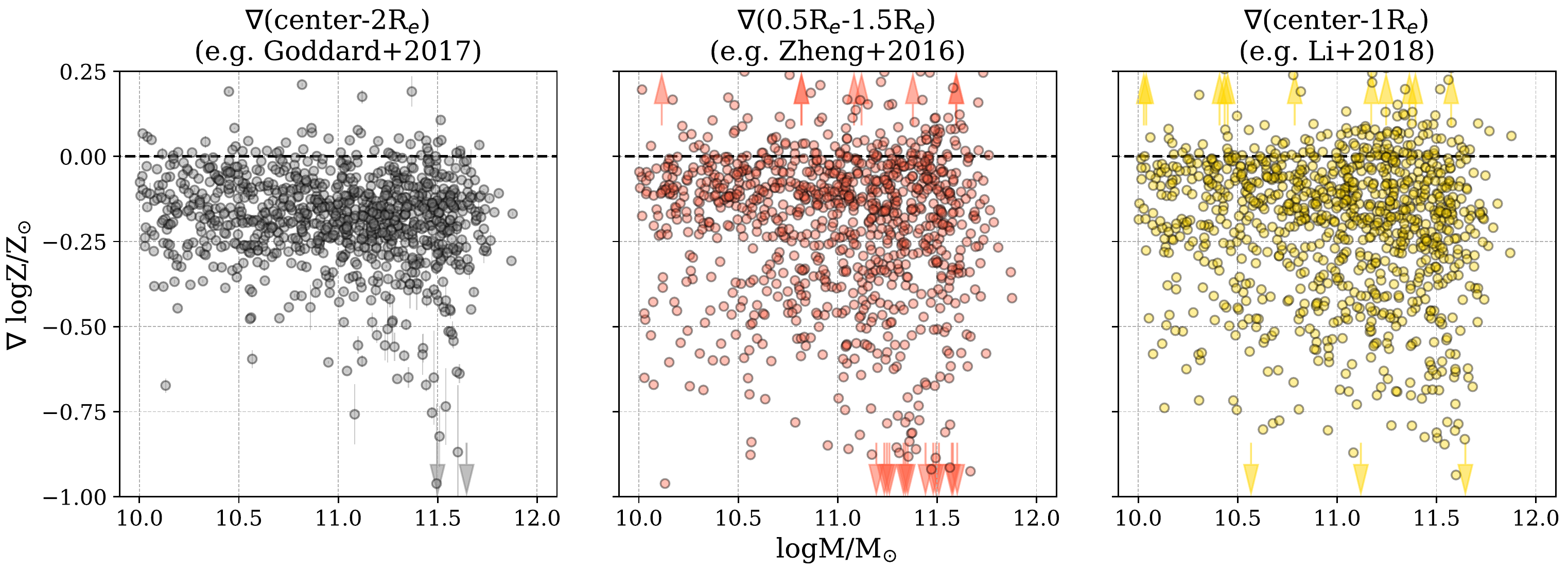}
		\caption{Radial metallicity gradients for our ETGs as a function of M$_*$. These gradients were computed by fitting a straight line to the radial profiles. From left to right, we fit the radial ranges R$<$2R$_{\mbox{\scriptsize e}}$, 0.5R$_{\mbox{\scriptsize e}}<$R$<$1.5R$_{\mbox{\scriptsize e}}$, and R$<$1R$_{\mbox{\scriptsize e}}$. The arrows indicate gradients beyond the scale of the figure. Note how gradients fail to capture most of the high M$_*$ flattening seen in Figure \ref{fig7}. This figure was made with the outputs from Prospector, but results stand for Firefly and pPXF.}
		\label{fig8}
	\end{figure*}

	A quick look at our Figure \ref{fig7} reveals that the average metallicity profiles of ETGs are not straight lines. It stands to reason that fitting lines to these radial profiles could ``wash-out" the flattening in the outskirts of high M$_*$ ETGs. Figure \ref{fig8} shows the outcome of fitting lines to our metallicity profiles over different radial ranges motivated by the literature. Some ranges trace the inner regions (R$<$R$_e$; \citealt{li2018}), while others have more extended coverage (R$<$2R$_e$; \citealt{goddard2017a}). The scatter is considerable in all cases, and recovering any correlations with M$_*$ is difficult. We conclude that gradients are sensitive to radial coverage (see also \citealt{greene2019}) and can also miss important behavior in the stellar metallicity profiles. Gradients should be avoided when possible.

	\begin{figure}
	\centering
	\includegraphics[width=3.4in]{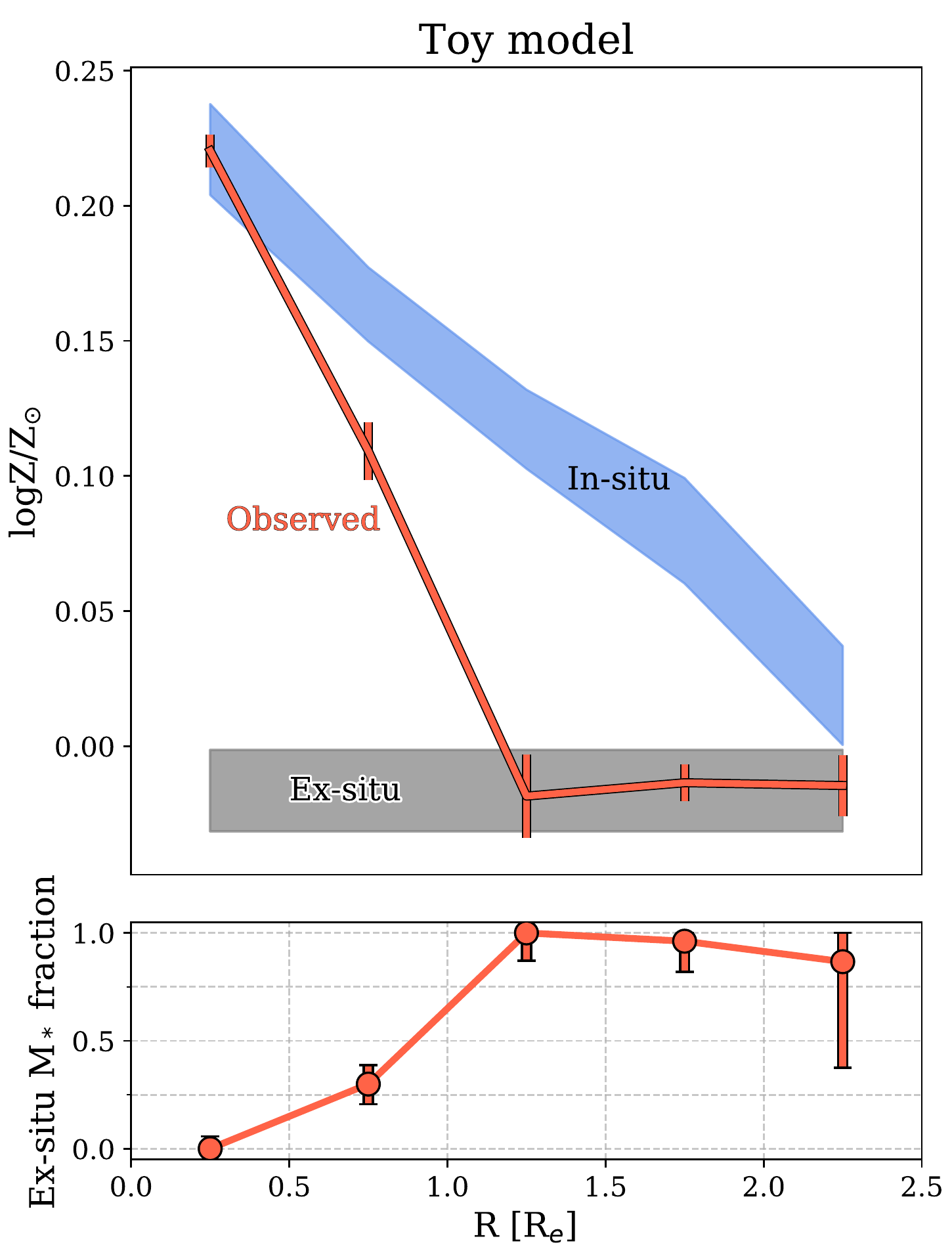}
	\caption{Top: Decomposition of the observed metallicity profile (red data points) in the highest M$_*$ bin ($10^{11.5}-10^{12}$M$_{\odot}$).  We ascribe the \textit{in-situ} component in this mass bin the same shape as the observed metallicity profile in the lowest M$_*$ bin ($10^{10}-10^{10.5}$M$_{\odot}$), but scaled upward to match the observed, central metallicity at higher M$_*$. The \textit{ex-situ} component (grey) is ascribed a single metallicity $\varepsilon\sim -0.24$ lower than the observed central metallicity.  The mix of components lowers the observed metallicity at all radii.  Bottom: The amount of suppression determines the required fraction of \textit{ex-situ} stars at each radius. This figure was made with the outputs from Prospector, but also applies to Firefly and pPXF.}
	\label{fig9}
\end{figure}

	\subsection{Comparison with hydrodynamical simulations}
	\label{5.2}		
		
	Hydrodynamical simulations predict stellar accretion to induce gradient flattening (e.g. \citealt{cook2016}). In general, stars accreted via dry, minor mergers tend to settle around and beyond the outskirts of ETGs (R=2-4R$_{\mbox{\scriptsize e}}$), which results in a flatter stellar metallicity profile than the inherently steeper form it originally had. Since mergers are expected to have a larger effect on more massive systems, this prediction is in broad agreement with our results from Figure \ref{fig7}.
	
	A relevant point involves the radii at which accretion signatures are expected to appear. \citet{rodriguez-gomez2016} derived the accreted mass fraction of galaxies as a function of galactocentric radius in the Illustris simulation (\citealt{vogelsberger2014a,vogelsberger2014b}). On average, this fraction increases with radius. It goes from zero at the center to unity at radii R$\gtrsim5$R$_{\mbox{\scriptsize e}}$. This motivates the definition of the transition radius (R$_{\mbox{\scriptsize T}}$). It is defined as the galactocentric radius at which the M$_*$ fraction of the \textit{ex-situ} stellar component overtakes its \textit{in-situ} counterpart (\citealt{dsouza2014}). \citet{rodriguez-gomez2016} found R$_{\mbox{\scriptsize T}}$ to decrease with M$_*$, going from R$_{\mbox{\scriptsize T}}\sim 5$R$_{\mbox{\scriptsize e}}$ at M$_*\sim10^{10}$M$_{\odot}$ to R$_{\mbox{\scriptsize T}}<$R$_{\mbox{\scriptsize e}}$ at M$_*\sim10^{12}$M$_{\odot}$. Our results are qualitatively consistent with this prediction. 
	
	However, there are some quantitative tensions. For logM$_*$/M$_{\odot}\sim 11$ galaxies, \citet{rodriguez-gomez2016} reported R$_{\mbox{\scriptsize T}}\sim4$ R$_{\mbox{\scriptsize e}}$. Within 2.5R$_{\mbox{\scriptsize e}}$, we should only be probing accreted stellar mass fractions of $\lesssim$0.3 at this mass range. \citet{cook2016}, also based on the Illustris simulation, reported that the flattening of metallicity gradients with M$_*$ only becomes noticeable in the stellar halo (R$=2-4$R$_{\mbox{\scriptsize e}}$). Therefore, the signatures we see in Figure \ref{fig7} are apparent at smaller radii than some simulations have predicted. There are a few possible explanations for this tension. The works of \citet{rodriguez-gomez2016,cook2016} were based on Illustris. Galaxies at $z\sim 0$ from the first generation of this simulation were found to be larger by a factor of $\sim 2$ than observed galaxies. IllustrisTNG solved this problem, among others, by improving the treatment of galactic winds, magnetic fields, and black hole feedback (\citealt{pillepich2018,weinberger2017,weinberger2018}). The treatment of these, among with other secular processes, can strongly impact the stellar population gradients measured in simulations (e.g. \citealt{taylor2017}). On the observational side, estimates of the ages and metallicities of stellar populations can strongly depend on the choice of stellar library, isochrones, and approach to fitting. These systematic uncertainties also affect the conversion between stellar mass and stellar light, impacting the comparison between simulations and observations.
	
	
	\begin{figure*}
		\centering
		\includegraphics[width=7in]{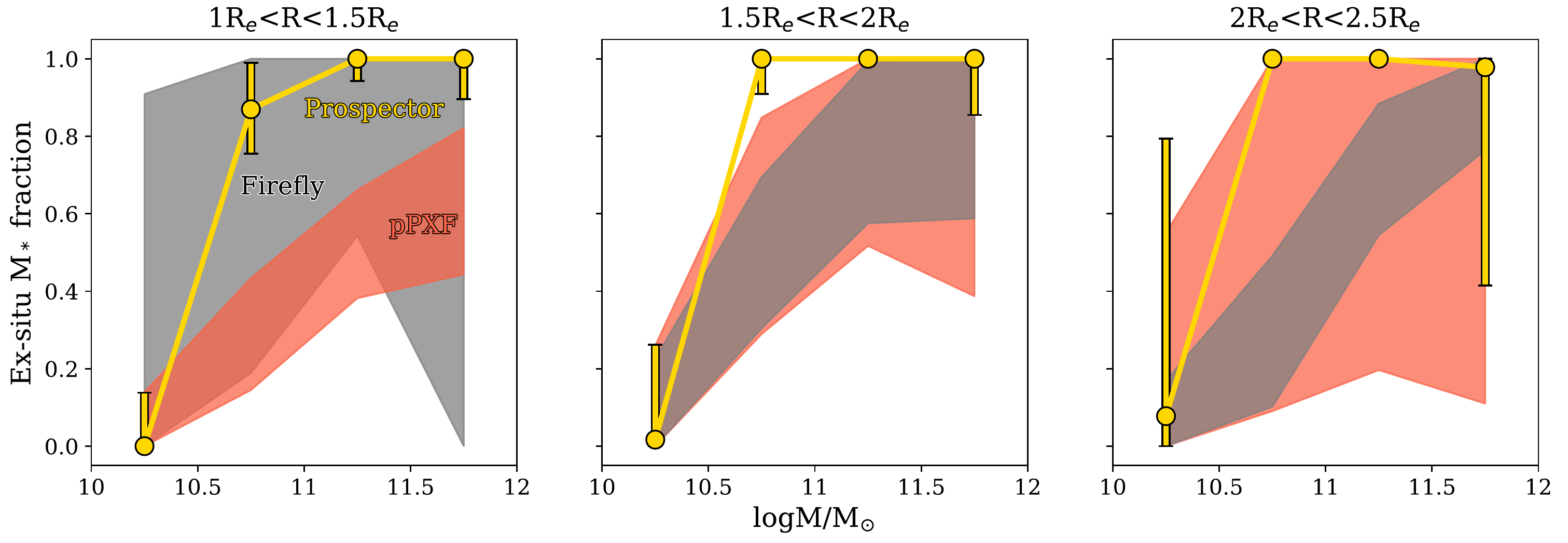}
		\caption{Observational estimate of the \textit{ex-situ} stellar mass fraction in ETGs as a function of M$_*$ for three different radial bins. Shown are the 1$\sigma$ contours derived with Firefly (grey), pPXF (red), and Prospector (yellow). The estimates come from expressing the metallicity profiles of ETGs as a linear combination of \textit{in-situ} and \textit{ex-situ} profiles (Figure \ref{fig9}). Note how \textit{ex-situ} signatures increase with M$_*$.}
		\label{fig10}
	\end{figure*}
	
	\subsection{Estimating the \textit{ex-situ} stellar mass fraction}
	\label{5.3}
	
	Observationally, global stellar metallicity correlates with M$_*$ or the central velocity dispersion $\sigma_*$ of galaxies (\citealt{faber-jackson1976,cidfernandes2005,gallazzi2005,thomas2005,thomas2010,gonzalez-delgado2014}), as would be expected if the deeper potential wells of more massive systems limit the impact of galactic winds (\citealt{matteucci1994}). In the Illustris simulation, \citet{dsouza-bell2018} found an accreted M$_{\mbox{\scriptsize acc}}$-Z$_{\mbox{\scriptsize acc}}$ relation, where M$_{\mbox{\scriptsize acc}}$ and Z$_{\mbox{\scriptsize acc}}$ refer to the stellar mass and stellar metallicity of the accreted components, respectively. This relationship lies $\sim 0.3$ dex below the global counterpart. We can make informed assumptions for the \textit{in-situ} stellar metallicity profile and the M$_{\mbox{\scriptsize acc}}$-Z$_{\mbox{\scriptsize acc}}$ relation to build a toy model capable of inferring the \textit{ex-situ} M$_*$ fraction as a function of mass and galactocentric radius from our observations.
		
	We assume the intrinsic \textit{in-situ} metallicity profiles of ETGs to be well described by the profiles observed in the low mass end of our sample. This is supported by hydrodynamical simulations that find M$_*\sim 10^{10}$M$_{\odot}$ galaxies to be dominated by \textit{in-situ} stars within the radial coverage of our data (\citealt{rodriguez-gomez2016}). We take the M$_*=10^{10}-10^{10.5}$M$_{\odot}$ profiles from Figure \ref{fig7} for each code and refer to them as
	\begin{gather}
   \mbox{logZ}_{\mbox{\scriptsize obs}}(\mbox{R, low M}_*)
	\end{gather} 	
	In our model, the \textit{in-situ} profiles of all galaxies follow the shape of $\mbox{logZ}_{\mbox{\scriptsize obs}}(\mbox{R, low M}_*)$ with a normalization applied to match the metallicity at the center (i.e. within 0.5 R$_{\mbox{\scriptsize e}}$). This can be written as
	\begin{gather}
	\label{toy}
	\mbox{logZ}_{\mbox{\scriptsize in-situ}}(\mbox{R, M}_*)=	\mbox{logZ}_{\mbox{\scriptsize obs}}(\mbox{R, low M}_*) \\
	\nonumber
	+\mbox{logZ}_{\mbox{\scriptsize obs}}(\mbox{0.25R$_{\mbox{\scriptsize e}}$, M}_*)-\mbox{logZ}_{\mbox{\scriptsize obs}}(\mbox{0.25R$_{\mbox{\scriptsize e}}$, low M}_*)
	\end{gather} 
	with a schematic representation in Figure \ref{fig9}.
	
	The M$_{\mbox{\scriptsize acc}}$-Z$_{\mbox{\scriptsize acc}}$ relation is offset 0.3 dex from the global counterpart in the Illustris simulation. The existence of this relation originates from single massive progenitors contributing to the bulk of the mass to the accreted stellar component (\citealt{dsouza-bell2018}). If we assume the accreted envelopes of ETGs to be comparable in stellar mass to their host ETG (e.g. \citealt{rodriguez-gomez2016}), \textit{ex-situ} metallicities can be approximated by
	\begin{gather}
	\label{toy1}
	\mbox{logZ}_{\mbox{\scriptsize ex-situ}}(\mbox{M}_*)=\mbox{logZ}_{\mbox{\scriptsize obs}}(\mbox{0.25R$_{\mbox{\scriptsize e}}$, M}_*) - \varepsilon
	\end{gather} 	
	i.e., stellar metallicity of \textit{ex-situ} stars will be $\varepsilon=0.3$ dex lower than the metallicity at the center of the galaxy. Note that there is no dependence on galactocentric radius in the definition of $\mbox{logZ}_{\mbox{\scriptsize ex}}$.
	
	For measured metallicities, the offset will be dependent on the stellar population synthesis approach. To account for differences between codes, we set $\varepsilon$ equal to the difference in metallicity between the centers and outskirts of M$_*=10^{11.5}-10^{12}$M$_{\odot}$ ETGs (see Figure \ref{fig9}). The corresponding values are $\varepsilon \sim -0.14$ (Firefly), $-0.29$ (pPXF), and $-0.24$ (Prospector).
	
	We can now write observed metallicities as a linear combination between \textit{in-situ} and \textit{ex-situ} metallicities:
	\begin{gather}
	\label{toy2}
		\nonumber
	\mbox{logZ}_{\mbox{\scriptsize obs}}(\mbox{R, M}_*)=f_{\mbox{\scriptsize in-situ}}(\mbox{R, M}_*)\mbox{ logZ}_{\mbox{\scriptsize in-situ}}(\mbox{R, M}_*) \\ 
	+f_{\mbox{\scriptsize ex-situ}}(\mbox{R, M}_*)\mbox{ logZ}_{\mbox{\scriptsize ex-situ}}(\mbox{M}_*)
	\end{gather} 	
	where $f_{\mbox{\scriptsize in-situ}}$ and $f_{\mbox{\scriptsize ex-situ}}=1-f_{\mbox{\scriptsize in-situ}}$ are the \textit{in-situ} and \textit{ex-situ} fractions. Figure \ref{fig9} describes our toy model and how we derive \textit{ex-situ} fractions from it.
	
	The results as a function of M$_*$, galactocentric radius, and code are shown in Figure \ref{fig10}. \textit{Ex-situ} fractions increasingly dominate at larger radii and higher M$_*$. \textit{Ex-situ} stars at R$\sim$2R$_{\mbox{\scriptsize e}}$ make up $\lesssim$20\% of the projected stellar mass of M$_*\lesssim 10^{10.5}$M$_{\odot}$ ETGs, rising up to $\gtrsim$80\% for M$_*\gtrsim 10^{11.5}$M$_{\odot}$ ETGs.

	Stellar accretion and minor mergers provide an explanation for the size growth of spheroids from $z\sim2$ to the present. Keeping in mind the simple nature of our comparison, we showed that the logZ profiles of nearby ETGs are consistent with this framework. However, this picture might not apply to S0s, which we visually estimate to compose $\lesssim 20$\% of our sample. The growth and accretion histories of S0s can differ from those of elliptical galaxies ($n_{\mbox{\scriptsize Sersic}} >2.5$; \citealt{blanton2003,blanton2005,peng2010}), as suggested by \citet{johnston2012,johnston2014,fraser-mckelvie2018,saha2018} (see also \citealt{diaz2018}). Moreover, galaxy assembly history is not only expected to depend on the total M$_*$ or morphology of galaxies, but also on their environment (e.g. \citealt{greene2015,greene2019}). We will study second order trends in the metallicity profiles of ETGs in follow-up work.
	
	\section{Summary}
	\label{6}
	
	We characterized the radial stellar metallicity profiles of MaNGA ETGs and compared them with predictions from hierarchical formation. Through stellar population fitting with Firefly, pPXF, and Prospector, we found the following:
	
	1. The three codes are built around different stellar population synthesis codes and are unique in their approach to fitting. Nonetheless, we found the main conclusions from this paper not to be dependent on the fitting code.
	
	2. The profiles of logM$_*$/M$_{\odot}\gtrsim 11$ ETGs fall with galactocentric radius and flatten beyond R$\sim$1.5R$_{\mbox{\scriptsize e}}$. Based on hydrodynamical simulations, a possible explanation for this flattening is stellar accretion through minor mergers. 
	
	3. The average radial metallicity profiles of ETGs are not linear. Therefore, linear fits can miss important behavior in the stellar metallicity profiles. When possible, fitting stellar population gradients should be avoided.
	
	4. Using informed assumptions for the \textit{in-situ} metallicity profile and the metallicity of accreted stars, we built a toy model to infer the \textit{ex-situ} stellar mass fraction of ETGs. We found \textit{ex-situ} signatures to grow in significance toward large galactocentric radii and higher M$_* $.
	
	\acknowledgments
	
	We thank the referee for their constructive comments and suggestions. We also thank Ben Johnson for helping us run Prospector on MaNGA data. We are grateful to Platon Karpov and Enrico Ram\'irez-Ruiz for allowing us to use their cluster, Comrade, to run stellar the stellar population fitting codes. This work made use of GNU Parallel (\citealt{tange2018}) for executing jobs on Comrade. We thank Brian Cherinka and Jos\'e S\'anchez-Gallego for helping the authors to familiarize themselves with Marvin, the core Python package and web framework for MaNGA data. GO acknowledges support from the Regents' Fellowship from the University of California, Santa Cruz. KB is supported by the UC-MEXUS-CONACYT Grant. ZZ is supported by the National Natural Science Foundation of China No. 11703036. This research made use of Marvin, a core Python package and web framework for MaNGA data, developed by Brian Cherinka, Jos\'e S\'anchez-Gallego, and Brett Andrews (MaNGA Collaboration, 2018). Funding for the Sloan Digital Sky Survey IV has been provided by the Alfred P. Sloan Foundation, the U.S. Department of Energy Office of Science, and the Participating Institutions. SDSS acknowledges support and resources from the Center for High-Performance Computing at the University of Utah. The SDSS web site is www.sdss.org. SDSS is managed by the Astrophysical Research Consortium for the Participating Institutions of the SDSS Collaboration including the Brazilian Participation Group, the Carnegie Institution for Science, Carnegie Mellon University, the Chilean Participation Group, the French Participation Group, Harvard-Smithsonian Center for Astrophysics, Instituto de Astrof\'isica de Canarias, The Johns Hopkins University, Kavli Institute for the Physics and Mathematics of the Universe (IPMU) / University of Tokyo, Lawrence Berkeley National Laboratory, Leibniz Institut für Astrophysik Potsdam (AIP), Max-Planck-Institut für Astronomie (MPIA Heidelberg), Max-Planck-Institut für Astrophysik (MPA Garching), Max-Planck-Institut für Extraterrestrische Physik (MPE), National Astronomical Observatories of China, New Mexico State University, New York University, University of Notre Dame, Observat\'orio Nacional / MCTI, The Ohio State University, Pennsylvania State University, Shanghai Astronomical Observatory, United Kingdom Participation Group, Universidad Nacional Aut\'onoma de M\'exico, University of Arizona, University of Colorado Boulder, University of Oxford, University of Portsmouth, University of Utah, University of Virginia, University of Washington, University of Wisconsin, Vanderbilt University, and Yale University.
	\\ 

	\appendix
	\section{Lick index profiles}
	\label{appendix1}
	
	Lick indices (\citealt{worthey1994,thomas2003,parikh2019}) are a useful method to empirically estimate the chemical abundance patterns of galaxies. Here, we compute the radial profiles of Mgb(5178), Fe5270, and Fe5335 to test the high M$_*$ flattening we find through stellar population fitting. We retrieved the indices measured by the MaNGA Data Analysis Pipeline and used them to compute:
	\begin{gather}
     \mbox{[MgFe]'}=\sqrt{\mbox{Mgb}(0.72\times \mbox{Fe5270}+0.28\times \mbox{Fe5335})} \\
     \mbox{$<$Fe$>$}=0.5\times \mbox{Fe5270}+0.5\times \mbox{Fe5335}
	\end{gather} 
	Here [MgFe]' and $<$Fe$>$ are tracers of the global and Iron abundances (\citealt{johnston2018}). We binned the measurements into the five annuli R/R$_{\mbox{\scriptsize e}}$= $[0, 0.5]$, $[0.5, 1]$, $[1, 1.5]$, $[1.5, 2]$, and $[2, 2.5]$ to derive the median profiles shown in Figure \ref{fig10}. Note how the profiles flatten for the highest M$_*$ bin.
	
	\begin{figure*}
		\centering
		\includegraphics[width=7in]{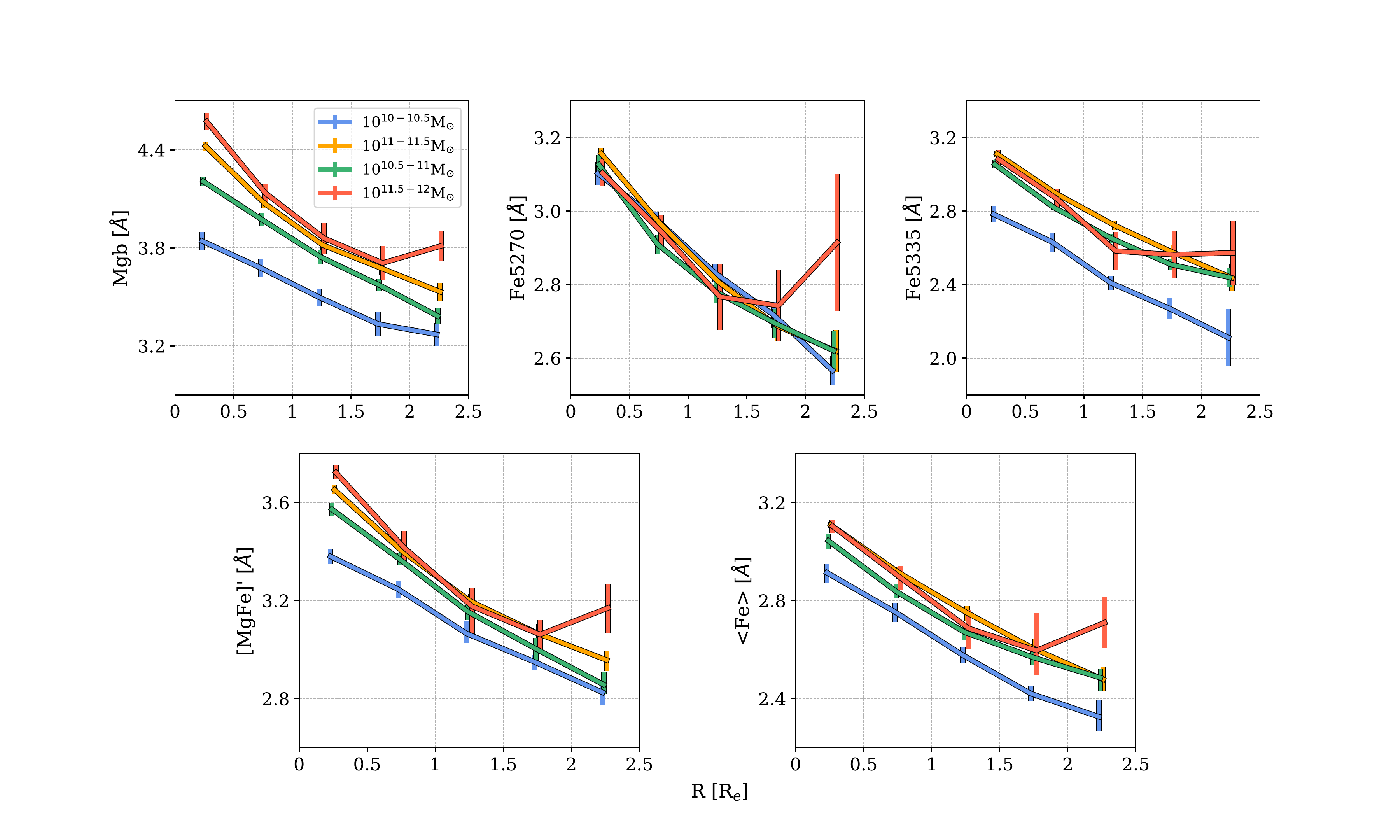}
		\caption{Median radial profiles of Mgb, Fe5270, Fe5335, [MgFe]', and $<$Fe$>$ for MaNGA ETGs as a function of M$_*$. These profiles are based on measurements made by the MaNGA Data Analysis Pipeline. The profiles flatten at the highest M$_*$, in consistency with Figure \ref{fig7}.}
		\label{fig11}
	\end{figure*}
	
	\bibliography{mybib}

	
\end{document}